\documentstyle[12pt]{article}
\title{ Glass state of the superfluid $^3$He-A in aerogel.}
\author{ G.E. Volovik\\
Low Temperature Laboratory, Helsinki University of Technology\\
Otakaari 3A, 02150 Espoo, Finland\\
and\\L.D. Landau Institute for Theoretical Physics, \\
Kosygin Str. 2, 117940 Moscow, Russia\\}
\begin{document}

\maketitle
\vskip 2 truecm

\begin{abstract}
{Glass states formed in the superfluid $^3$He confined in aerogel are
discussed. If the short range order corresponds to the A-phase state, the
glass state is nonsuperfluid in the long wave length limit. The superfluidity
can be restored by application of a small mass current. Transitions between
the superfluid and nonsuperfluid glass states can be triggered by small
magnetic field and by the change of the tipping angle of magnetization in NMR
experiments.}
\end{abstract}
\vskip 2 truecm

Submitted to JETP Letters, January 26, 1996

\vfill
\eject

In aerogel there is 98\% of the empty volume,  which is occupied by $^3$He,
while the other 2\% has mainly the shape of tangled silica strings or
strands. The string radius is about $R=20-30~ A^\circ$,  and the  average
distance between them is about $l\sim 1000  A^\circ$ \cite{Porto,Sprague}.
Since $R$ is much less than the superfluid coherence length $\xi_0$,  which
changes in the range  200-800 $ A^o$ depending on the pressure, these strings
have an effect of anisotropic and   randomly  oriented impurities. In the
triplet $p$-wave condensate such impurities have several different effects:
(i) They suppress both the condensate and the superfluid transition
temperature $T_c$, which is experimentally measured \cite{Porto,Sprague}. (ii)
They produce the renormalization of the coefficients in fourth order terms in
the Ginzburg-Landau functional, which according to \cite{GorkovKalugin}
triggers the transition between different superfluid phases in the anisotropic
heavy fermionic superconductors. The same arguments applied to the $^3$He in
aerogel \cite{Thuneberg} explain why instead of the B-phase state in the bulk
superfluid $^3$He the magnetic experiments \cite{Sprague} provide  an evidence
of the A-phase (or of another equal pairing state). (iii) Also the  random
local anisotropy produced by the string network leads according to   Imry  and
Ma \cite{ImryMa}  to the violation of the orientational long range order at
large distances. This was discussed  for anisotropic superconductors in
\cite{VolKhmel}, where the state without the long range orientational order was
called the superconducting glass.

Here we discuss some properties of the "superfluid glass" states of the
$^3$He in the aerogel. In the case of the exotic superconductivity described
by the vector order parameter the effect of the random
anisotropy is suppressed by the regular anisotropy of the crystal lattice, as
a result  the superconducting glass can arise only in a small strip of the
phase diagram very close to $T_c$  \cite{VolKhmel}. In   superfluid $^3$He
the external regular anisotropy is absent that is why  the  superfluid $^3$He
in aerogel is always in the glass state in the absence of
external magnetic field. Application of  rather weak magnetic field can
restore the conventional long range order.

The main property of the glass state is that the expectation value of the order
parameter (averaged over the large enough volume) vanishes:  $<A_{\alpha
i}(\vec r)>=0$. In other words the correlation of the order
parameter decays at large distances: $<A_{\alpha i}(\vec r_1) A_{\beta
j}(\vec r_2)>=0$ if $(\vec r_1 -\vec r_2)/L_0 \rightarrow \infty$. Here $L_0$
is the characteristic length, within which the order parameter is
homogeneous. The value of $L_0$ will be discussed below.  Some part
of the long range order can still be remained and it is described in terms of
bilinear combinations, such as
$$Q_{\alpha  \beta ij}=<A_{\alpha i}(\vec r) A^*_{\beta j}(\vec
r)>~~, \eqno(1)$$   or
$$P_{\alpha  \beta ij}=<A_{\alpha i}(\vec r) A_{\beta j}(\vec
r)>~~.\eqno(2)$$

Let us assume that the short range order corresponds to the A-phase, which is
consistent with the magnetic measurements \cite{Sprague}. The local order
parameter of the A-phase is $A_{\alpha i}(\vec r)=\Delta d_{\alpha}(e_{1
i}(\vec r)+ie_{2 i}(\vec r))$ \cite{VolWol}, where $\hat d$ is the unit vector
of spin anisotropy, and the orthogonal unit vectors $\hat e_1$ and $\hat e_2$
describe the orbital part of the order parameter, which is randomly oriented
in space. Assuming that silica strands violate only the orbital orientational
order, one has
$$P=0~~,~~Q_{\alpha  \beta ij}=Q d_{\alpha}  d_{\beta}
\delta_{ij}~~.\eqno(3)$$
This long range order corresponds to the nonsuperfluid spin nematic.
The  quantity $Q$ is gauge invariant and thus the state with $Q\neq 0$, $P=0$
is nonsuperfluid.

The difference between the superfluid and nonsuperfluid
states is given by the loop function \cite{Kosterlitz}:
$$<e^{i(2\pi/\kappa)\oint_L \vec v_s\cdot d\vec r} >~~, $$
where  $\kappa=\pi \hbar/m_3$ is the circulation quantum in $^3$He and
integral is along the loop of length  $L$.  In superfluids the loop function
decays as $e^{-L}$ or slower, while in the nonsuperfluid state it decays as
$e^{-S}$, where $S \sim L^2$  is the area of the loop. Using the  Mermin-Ho
relation \cite{Mermin-Ho}, which couples the continuous vorticity of
superflow with the spatial variation of the randomly oriented orbital
anisotropy vector
$\hat l=
\hat e_1\times\hat e_2$:
$$
\vec  \nabla\times\vec  v_s =
      {\kappa\over 4}e_{ijk} \hat l_i\vec  \nabla \hat l_j\times\vec
\nabla
\hat l_k ~~,
\eqno(4)
$$
one finds the area law for the state in Eq.(3):
$$<e^{i(2\pi/\kappa)\oint_L \vec v_s\cdot d\vec r }>~ \propto
e^{-L^2/L_0^2} ~~,~~L \gg L_0~~.\eqno(5)$$

The Eq.(5) means that the linear response of the current to the superfluid
velocity $\vec v_s$ vanishes in the long wave length limit: $\rho_s (q \ll
1/L_0)=0$. The superfluid density is however restored to its local value  if
the velocity is large enough:  $v_s \gg \kappa/L_0$. This can be directly
checked in the experiments of the type described in \cite{Porto}, if one
measures $\rho_s$ as a function of $v_s$. This type of behavior can be named
"weak nonsuperfluidity".

Till now we ignored the tiny spin-orbit interaction $-g_D(\hat l\cdot \hat
d)^2$ between the spin and orbital degrees of freedom, which is characterized
by the so called dipole length $\xi_D \sim 10^{-3}$cm \cite{VolWol}. This is
justified if $L_0 < \xi_D$. If $L_0 > \xi_D$,   the vector $\hat d$  is
"dipole-locked" with $\hat l$ and also becomes randomly oriented. In
this case $Q_{\alpha  \beta ij}$ becomes isotropic and at large distances the
symmetry of the state corresponds to that of the normal $^3$He without any
long range order.

Situation changes if the magnetic field $H \gg 30$Gauss is
applied. In this case the spin part of the order parameter is oriented
perpendicular to magnetic field: $\hat d \perp \vec H$. Since  $L_0 > \xi_D$
the $\hat l$ vector is to be aligned with $\hat d$. The order parameter
$P_{\alpha  \beta ij}$ in Eq.(2) now contains  nonzero terms,  one of them is:
$$P_{\alpha  \beta ij}=P(\delta_{\alpha \beta}-\hat z_\alpha \hat
z_{\beta} )(\delta_{ij} - 3\hat z_i \hat z_j)~e^{2i\Phi}~~.\eqno(6)$$
This  state with $P_{\alpha  \beta ij}\neq 0$ is equivalent to the superfluid
condensate, the elementary boson of which is  not a Cooper pair of atoms but
rather consists of 4 atoms. Thus the transition from the weak nonsuperfluid to
the superfluid states can be triggered by rather small magnetic field of order
100 G.

Another possible way to trigger the transition is to change the dipole
energy. This can be made in the regime of nonlinear NMR with the large tipping
angle $\beta$ of the precessing magnetization. According to \cite{Bunkov}
the dipole energy which tends to keep the vector $\hat l$ in the plane
perpendicular to $\vec H =H\hat h$ is
$$F_D=  g_D    ({7\over 8}\cos^2\beta +
{1\over 4}\cos\beta -{1\over 8}) (\hat l \cdot \hat h)^2 ~~.\eqno(7)$$
The dipole energy decreases with increasing $\beta$ and at some moment
$\xi_D(\beta)= \xi_D(0) ({7\over 8}\cos^2\beta +
{1\over 4}\cos\beta -{1\over 8})^{-1/2}$ becomes larger than $L_0$. The
critical value   $\beta_c$ at which the transition should occur is smaller than
the value of $\beta$ at which the dipole energy changes sign: $F_D=0$ at
$\cos\beta =0.26$ and thus $\beta_c < 75^\circ$. This correlates with the
abrupt change of the NMR signal observed in \cite{Sprague} at $\beta_c
\sim 40^\circ -50^\circ$. The details of this transition are however
unclear.

Now let us discuss the magnitude of the correlation length $L_0$ at which the
random anisotropy  produced by anisotropic silica strands  kills
the orientational  and superfluid orders in $^3$He-A. Aerogel can be
represented as the randomly distributed cylinders  of the radius $R$ with the
distance $l$ between the cylinders. According to
\cite{RainerVuorio} the characteristic energy of the order parameter
distortion produced by a  small object of size $R\ll \xi_0$   is
$$k_F^2 R   \Delta^2/T_c  ~~\eqno(8)$$
per unit length of the strand.  This is the measure of the anisotropy energy
which tends to orient the $\hat l$-vector. Now let us consider the box of
size $L\gg l$  and apply Imry-Ma arguments \cite{ImryMa}.
The orientational energy, which comes from the typical
fluctuation within this box of volume $V=L^3$  and which tends to fix the
orientation of $\hat l$ in the box, is
$$E_{random}=(L/l)^{3/2} k_F^2 R l\Delta^2/T_c .\eqno(9)$$
The gradient energy which arises due to different preferred orientation
of $\hat l$ in neighbouring boxes is
$$E_{gradient}\sim K \int_V (\nabla \hat l)^2
\sim (\Delta/T_c)^2 (k_F^3/m) L   .\eqno(10)$$
These energies are equal at
$$L=L_0\sim l {\xi_0^2 \over R^2} \gg l  ,\eqno(11)$$
and this gives the size $L_0$  of the domain
within which the $\hat l$-vector is homogeneous. In this model $L_0$ is
temperature independent, while its value can be comparable with $\xi_D$, since
$R \ll \xi_0 < l$.

The value of $L_0$ can be found   by  measurement  of the linear
term  in the heat capacity of $^3$He in aerogel at low temperature:
$C(T)\propto N(0)T$. The nonzero density of  states $N(0)$ arises in the
$\hat l$ texture due to the gap nodes in $^3$He-A\cite{Exotic}:
$$N(0)\sim N_F \xi_0 \vert\hat l\times (\vec\nabla\times\hat l)\vert \sim
N_F {\xi_0\over L_0} \sim N_F {R^2\over \xi_0 l}~~,\eqno(12)$$
where $N_F$ is the density of states in the normal Fermi liquid.
The observation of this effect will be interesting, since it will be
the indication  of the anomalies in $^3$He-A, related to the point gap nodes,
which are very similar to anomalies in particle physics \cite{Exotic}.

I thank W.P. Halperin, Yu.G. Makhlin and E.V. Thuneberg for illuminating
discussions. This work was supported through the ROTA co-operation plan of the
Finnish Academy and the Russian Academy of Sciences and by the Russian
Foundation for Fundamental Sciences, Grant Nos. 93-02-02687 and 94-02-03121.

\vfill\eject

\end{document}